\documentclass[12pt,preprint]{aastex}

\usepackage{CJK}
\usepackage{natbib}
\usepackage{graphicx}
\usepackage{rotating} 
\usepackage{epsfig}

\begin{document}
\begin{CJK}{UTF8}{gbsn}

\title{Discovery of Molecular Hydrogen in White Dwarf Atmospheres}

\author{S. Xu(许\CJKfamily{bsmi}偲\CJKfamily{gbsn}艺)\altaffilmark{a}, M. Jura\altaffilmark{a}, D. Koester\altaffilmark{b}, B. Klein\altaffilmark{a}, B. Zuckerman\altaffilmark{a}}
\altaffiltext{a}{Department of Physics and Astronomy, University of California, Los Angeles CA 90095-1562; sxu@astro.ucla.edu, jura@astro.ucla.edu, kleinb@astro.ucla.edu, ben@astro.ucla.edu}
\altaffiltext{b}{Institut fur Theoretische Physik und Astrophysik, University of Kiel, 24098 Kiel, Germany; koester@astrophysik.uni-kiel.de}

\begin{abstract}
With the Cosmic Origins Spectrograph onboard the {\it Hubble Space Telescope}, we have detected molecular hydrogen in the atmospheres of three white dwarfs with effective temperatures below 14,000 K, G29-38, GD 133 and GD 31. This discovery provides new independent constraints on the stellar temperature and surface gravity of white dwarfs.
\end{abstract}


\section{INTRODUCTION}

An important discovery from the {\it International Ultraviolet Explorer (IUE)} was two broad absorption features at 1400 {\AA} and 1600 {\AA} in hydrogen-dominated (DA) white dwarfs cooler than 20,000 K and 13,500 K, respectively \citep{GreensteinOke1979, Wegner1982}. Subsequently, these features were explained by \citet{Koester1985} and \citet{NelanWegner1985} as Ly $\alpha$ satellite lines from collisions among H-H (1600 {\AA}) and H-H$^+$ (1400 {\AA}). These quasi-molecular absorption lines help to constrain stellar temperature and surface gravity for DA white dwarfs \citep{AllardKoester1992}. G29-38, GD 133 and GD 31, the targets described in this letter, all show these quasi-molecular absorption features when observed with the {\it IUE} [\citet{Holm1985, KeplerNelan1993}, {\it IUE} archive]. Here, we report first detections of true molecular hydrogen in these three white dwarfs, by employing the Cosmic Origins Spectrograph (COS) on the {\it Hubble Space Telescope (HST)}. These three stellar atmospheres are among the hottest stellar environments where photospheric molecular hydrogen has ever been detected.

\section{DATA}

G29-38 and GD 133 were observed as part of program 12290 during {\it HST} cycle 18, which focuses on using externally polluted white dwarfs to assess the volatile abundances of accreted extrasolar planetesimals [see \citet{Jura2003, Zuckerman2007, Klein2010, Dufour2012, Jura2012, Gaensicke2012} and references within]. G29-38 and GD 133 were chosen because they show a high degree of atmospheric pollution from optical studies \citep{Koester1997, Koester2005} and display excess infrared radiation \citep{ZuckermanBecklin1987, Reach2005b, Reach2009, Jura2007b}. As part of our program to study Hyades white dwarfs \citep{Zuckerman2013}, ultraviolet data for GD 31 were retrieved from the {\it HST} archive of the SNAPSHOT program 12169 (PI: B. G{\"a}nsicke).

For G29-38 and GD 133, the COS set-up was similar to that employed for GD 40 and G241-6, as previously reported in \citet{Jura2012}. The G130M grating was used with central wavelength 1300 {\AA} and wavelength coverage of 1142 -1288 {\AA} (strip B) and 1298 -1443 {\AA} (strip A).  The spectral resolution was $\sim$ 20,000. Total exposure times, from combining 4 separate exposures for G29-38 and 5 for GD 133, were 9032 sec and 13,460 sec, respectively. The signal-to-noise ratio is at least 8 for strip B and 15 for strip A for both stars. The raw data were processed under the pipeline CALCOS 2.18.5. Because G29-38 is a ZZ Ceti star with multiple pulsation modes \citep{Thompson2008}, which lead to considerable changes in flux levels and continuum shape in each individual exposure, CALCOS processing cannot fully correct for these factors. Instead, we fitted the continuum of each CALCOS-extracted spectrum with a low-order spline3 polynomial and combined the normalized spectra with IRAF. We did not find any noticeable differences in line strength in each exposure. GD 133 is also a ZZ Ceti but the pulsation amplitude is much lower \citep{Silvotti2006} and we simply adopt the pipeline reduced spectra. For GD 31, the G130M configuration has a central wavelength of 1291 {\AA}, as described in \citet{Gaensicke2012}. The total exposure time was 1200 sec and the signal-to-noise ratio is $\sim$ 7 for strip A. We also used the output from the CALCOS pipeline for GD 31.

Apart from the absorption lines from heavy elements in G29-38 and GD 133, which will be discussed in \citet{Xu2013b}, we see numerous quasi-periodic features, as illustrated in Figure 1. They are ubiquitous between 1310 {\AA} and 1443 {\AA}, almost the entire strip A. Clearly, these features are real and not instrumental artifacts, such as fixed pattern noise, for the following reasons. Unlike GD 40 and G241-6 \citep{Jura2012}, two other white dwarfs observed in the same program with the same instrumental set-up, these features were only present in the spectra of G29-38 and GD 133. Furthermore, they are at the same wavelength in each separate exposure. Although initially puzzled, after comparing with the ultraviolet spectrum of the Sun \citep{Sandlin1986}, we realized the absorption wavelengths exactly match those of molecular hydrogen. All the recognizable features correspond to Lyman band transitions from $\nu$$''$= 2, 3, 4, 5 to $\nu$$'$ = 0 and Werner band transitions from $\nu$$''$=2, 3 to $\nu$$'$ = 0 \citep{Abgrall1993a, Abgrall1993b}. It turns out that we have accidentally found H$_2$ in white dwarf atmospheres! We present the strongest Lyman band H$_2$ lines in Table \ref{Tab: H2}; they are all appreciably stronger than the Werner bands in our observed wavelength interval. 

We serendipitously identified molecular hydrogen in a third star, GD 31, which may be a high-mass escaping member of the Hyades cluster \citep{Zuckerman2013}. The data are noisier but clearly four absorption features are seen; these correspond to the strongest H$_2$ Lyman band lines and their blends in G29-38, at the correct wavelengths, as presented in Table \ref{Tab: H2} and Figure 1. 

\section{DISCUSSION}

According to the Saha equation, molecular hydrogen is most concentrated in low temperature, high density environments. Previously, ro-vibrational lines from H$_2$ have been detected in cool stellar atmospheres in the infrared \citep{Spinrad1964} and in the Sun's ultraviolet emission spectrum \citep{Jordan1978,Sandlin1986}. Our ultraviolet detection of photospheric H$_2$ introduces a new aspect to stellar physics. With stellar parameters of T$_*$ = 11,820 K, log g=8.40 for G29-38 \citep{Xu2013b} and T$_*$=12,121 K, log g = 8.005 for GD 133 \citep{Koester2009b}, we computed white dwarf model atmospheres \citep{Koester2010} and calculated the number density of molecular hydrogen relative to atomic hydrogen, n(H$_2$)/n(H), as shown in Figure 2. At maximum, molecular hydrogen is still 10$^{-4.7}$ and 10$^{-5.1}$ less than the amount of atomic hydrogen in G29-38 and GD 133, respectively. These computed ratios are comparable to, but at the higher end of the concentration of trace elements in heavily polluted white dwarfs \citep{Jura2012}. Because molecular hydrogen is distributed over a large number of ro-vibrational levels, each individual line is relatively weak. As discussed in \citet{Zuckerman2013}, H$_2$ is used to resolve a major temperature puzzle for GD 31. Due to its high gravity, the equivalent widths (EWs) of 4 detected H$_2$ lines in GD 31 (T=13,700 K, log g=8.67) are comparable to those in GD 133; there is a substantial amount of molecular hydrogen in the atmosphere of GD 31.

As shown in Figure 1, we computed the model spectra for G29-38 and GD 133 following \citet{Koester2010} with all the line data obtained from the Kurucz webpage\footnote{http://kurucz.harvard.edu/linelists.html} \citep{Kurucz1995} and the H$_2$ partition function from \citet{Irwin1981}. In a statistical sense, the model well represents the data and reproduces most molecular hydrogen features. However, the fit to individual lines is rather poor due to the lack of accurate broadening parameters. The Z Astrophysical Plasma Properties (ZAPP) Collaboration is actively involved in creating white dwarf photospheres in the lab and determining line broadening parameters \citep{Falcon2012}. Our understanding of the COS ultraviolet observations can be substantially improved with accurate physical parameters for the H$_2$ lines.

Another aspect of this discovery is that by constraining the abundance of the HD molecule, we can place a limit on the D/H ratio. It is generally believed that all deuterium is primordial from the Big Bang nucleosynthesis and destroyed in stellar interiors \citep{Epstein1976}. Any amount, if present, in the white dwarf atmosphere must come from some external source, likely relic planetesimals. The values are not very constraining for the targets presented here. But if cooler stars can be observed with COS with a sufficiently high signal-to-noise ratio, then more meaningful upper limits or actual detections of the HD molecule in white dwarf atmospheres may be anticipated.

In white dwarfs cooler than 12,000 K, H$_2$ is present but usually there is not enough ultraviolet flux to make observations in a time-efficient manner. The environment in hotter stars is typically more hostile for molecular hydrogen but high pressure may still enable detection of H$_2$.

\section{CONCLUSIONS}

With the {\it HST}, molecular hydrogen was detected for the first time in white dwarf atmospheres. The three stars, G29-38, GD 133 and GD 31, have temperatures between 11,800 K and 13,700 K. H$_2$ can be used as an independent constraint to white dwarf atmospheric conditions. This opens a door to many future explorations.

Support for program \# 12290 was provided by NASA through a grant from the Space Telescope Science Institute, which is operated by the Association of Universities for Research in Astronomy, Inc., under NASA contract NAS 5-26555. We also utilized observations obtained [from the Data Archive] at the Space Telescope Science Institute, which is operated by the Association of Universities for Research in Astronomy, Inc., under NASA contract NAS 5-26555. These observations are associated with program \# 12169. This work also has been partly supported by NSF grants to UCLA to study polluted white dwarfs.

\end{CJK}

\newpage
\bibliographystyle{apj}

\newpage
\begin{figure}
\centerline{
\epsfig{figure=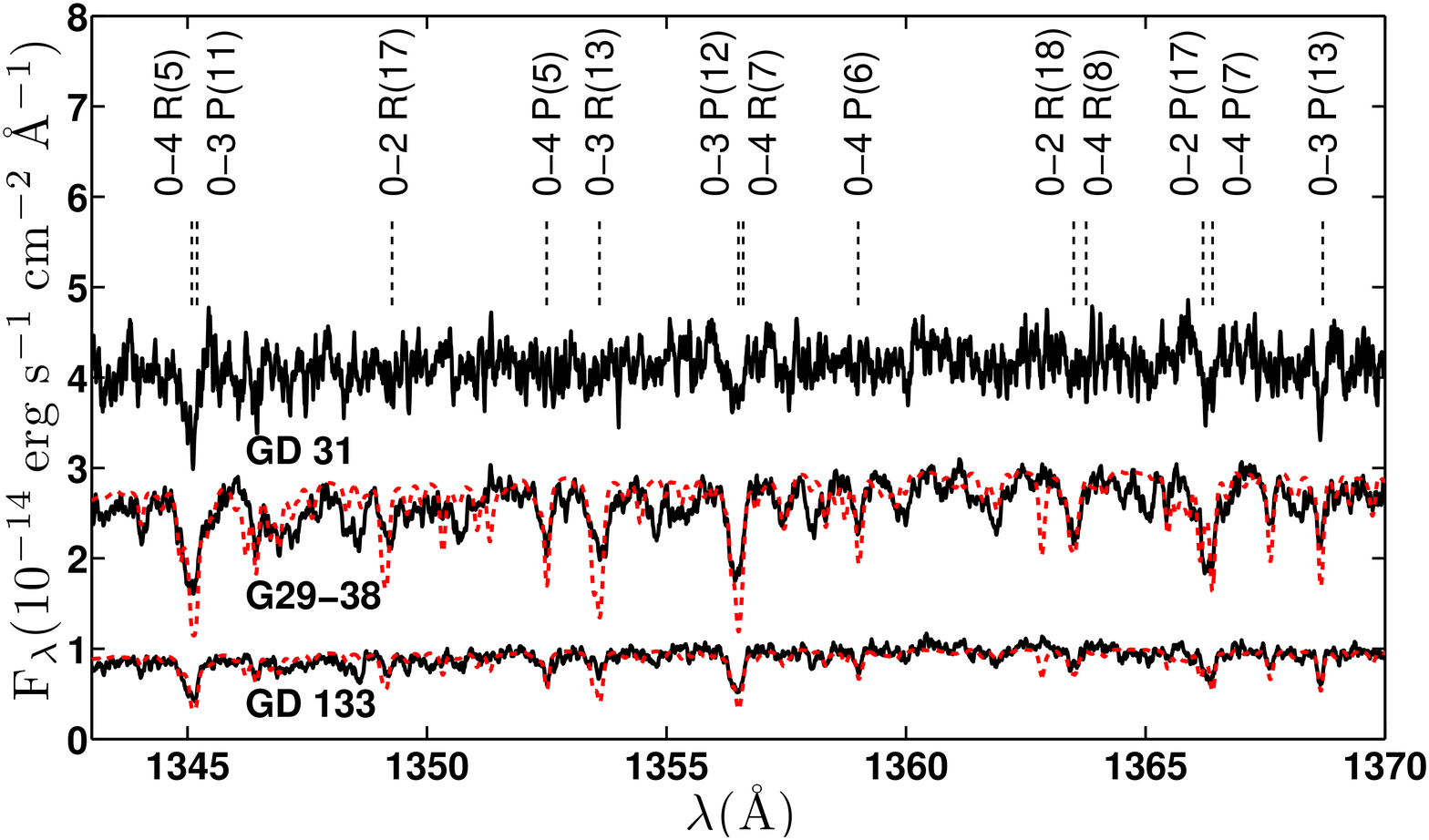,height=0.5\textheight,angle=90,clip=}
}
\end{figure}
\clearpage
\newpage
Fig 1. Portion of the Lyman band absorptions of molecular hydrogen in G29-38, GD 133 and GD 31 at the red wing of the H-H$^+$ quasi-molecular feature. The spectra are smoothed with a 5 point boxcar and shifted by 39 km s$^{-1}$, 58 km s$^{-1}$ and 89 km s$^{-1}$ for G29-38, GD 133 and GD 31 respectively, to be in the heliocentric reference frame \citep{Xu2013b,Zuckerman2013}. For clarity, the spectrum for GD 133 is offset by 0.5 $\times$ 10$^{-14}$ erg s$^{-1}$ cm$^{-2}$ {\AA}$^{-1}$ and for GD 31 by 1.5 $\times$ 10$^{-14}$ erg s$^{-1}$ cm$^{-2}$ {\AA}$^{-1}$. The red dashed lines are our computed model spectra. Statistically, the models reproduce the data but the fit to individual lines is not ideal. For G29-38 and GD 133, the entire spectrum in strip A (1298-1443 {\AA}) shows numerous ro-vibrational lines of H$_2$. The strongest absorption features around 1345 {\AA}, 1357 {\AA}, 1366 {\AA} and 1369 {\AA} are also seen in GD 31.

\newpage
\begin{figure}
\centerline{
\epsfig{figure=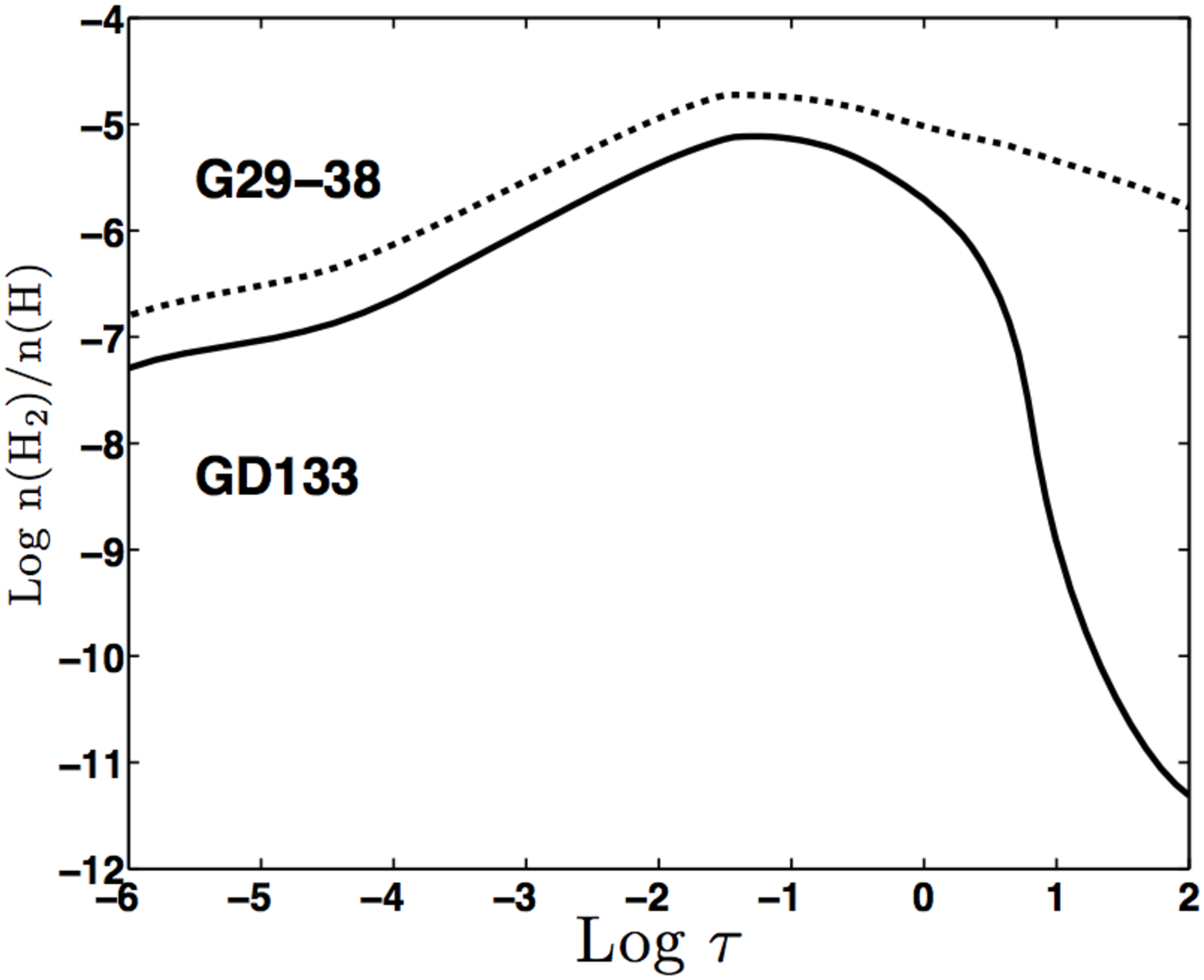,width=\textwidth,angle=0,clip=}
}
Fig 2. Ratio of number densities of molecular hydrogen compared to atomic hydrogen at different optical depths in G29-38 and GD 133. For G29-38, this ratio peaks at log n(H$_2$)/n(H) = -4.7, where $\tau$ = 0.04, T = 9900 K and $\rho$ =  3.1 $\times$ 10$^{-7}$ g cm$^{-3}$. For GD 133, the maximum of log n(H$_2$)/n(H) is -5.1, which occurs at $\tau$ = 0.05, T=10,200 K and $\rho$ = 1.5 $\times$ 10$^{-7}$ g cm$^{-3}$.
\end{figure}

\clearpage
\newpage
\begin{table}[hp]
\begin{center}
\caption{The Strongest Identified Lyman Band H$_2$ Lines}
\begin{tabular}{cclccccc}
\\
\hline \hline
	$\lambda$ & $\lambda$	& Transition$^b$& G29-38 & GD 133	& GD 31\\
  (cm$^{-1}$)$^a$	& ({\AA})& 	& EW (m\AA)	& EW (m\AA) & EW (m\AA)\\
  \hline
  74344.49	& 1345.1	& 0-4 R(5)	& 170$^c$ & 107$^c$	& 87$^c$	 	\\
  74339.87	& 1345.2	& 0-3 P(11)	& 170$^c$ & 107$^c$	& 87$^c$\\
  73936.94	& 1352.5	& 0-4 P(5)	& 76	&  66		& ...\\
  73719.80	& 1356.5	& 0-4 R(7)	& 121$^c$ & 103$^c$  & 63$^c$ \\
  73714.96	& 1356.6	& 0-3 P(12)	& 121$^c$ & 103$^c$ & 63$^c$ \\
  73185.32	& 1366.4	& 0-4 P(7)	& 180 & 70	& 59	 \\
  73064.05	& 1368.7	& 0-3 P(13)	& 45	& 35	& 35 \\
  71225.99	& 1404.0	& 0-4 P(11)	& 76 & 45  & ... \\
  71187.04		& 1404.8	& 0-5 R(5)	& 97 & 83	 & ...\\
  70992.76	& 1408.6	& 0-3 P(16) & 112$^c$ & 90$^c$ & ...\\
  70889.48	& 1410.6	& 0-4 R(13)	& 112$^c$ & 90$^c$	 & ...\\
  70076.46	& 1427.0	& 0-4 P(13)	& 75 & 53 & ...	\\
\hline
\label{Tab: H2}
\end{tabular}
\end{center}
\end{table}
\noindent $^a$ From \citet{Abgrall1993a}. \\
$^b$ We follow the notational conventions in \citet{Meyer2001}.\\
$^c$ This line is blended and the reported number is the total EW for both lines together. \\
{\bf Note.} For G29-38 and GD 133, the EW uncertainty is dominated by the choice of continuum interval because the whole region shows numerous H$_2$ lines and there is no clean continuum as shown in Figure 1. Typically, the EW uncertainty is within 10\%. For GD 31, the data are much noisier and the EW uncertainty is typically 25 \%.

\end{document}